\documentclass[11pt,reqno,a4paper]{amsart}
\usepackage{comment}
\usepackage{MnSymbol}              
\usepackage{graphicx}
\usepackage{color}
\usepackage{xcolor}
\usepackage{hyperref}
\usepackage{tikz}
\usetikzlibrary{patterns}
\usepackage{caption}
\usepackage{subcaption}
\hypersetup{colorlinks=true}
\numberwithin{equation}{section}
\numberwithin{figure}{section}
\newcommand{\beqn}{\begin{equation}}
\newcommand{\eeqn}{\end{equation}}
\newcommand{\beqna}{\begin{eqnarray}}
\newcommand{\ee}{\end{eqnarray}}
\newcommand{\best}{\begin{eqnarray*}}
\newcommand{\ees}{\end{eqnarray*}}
\newcommand{\bal}{\begin{align}}
\newcommand{\eal}{\end{align}}
\newcommand{\mb}{\mathbb}
\newcommand{\bp}{\begin{pmatrix}}
\newcommand{\ep}{\end{pmatrix}}
\newcommand{\al}{\alpha}
\newcommand{\om}{\omega}
\newcommand{\de}{\delta}

\newcommand{\be}{\beta}
\newcommand{\ga}{\gamma}

\newcommand{\e}{\epsilon}

\newcommand{\la}{\lambda}

\newcommand{\ol}[1]{\bar{#1}}
\newcommand{\oll}[1]{\bar{\bar{{#1}}}}
\newcommand{\oh}[1]{\hat{#1}}

\newcommand{\ohl}[1]{\hat{\bar{{#1}}}}
\newcommand{\ohhl}[1]{\hat{\hat{\bar{{#1}}}}}

\newcommand{\ot}[1]{\tilde{#1}}
\newcommand{\oht}[1]{\hat{\tilde{{#1}}}}
\newcommand{\olt}[1]{\bar{\tilde{{#1}}}}
\newcommand{\ohlt}[1]{\hat{\bar{\tilde{{#1}}}}}

\newcommand{\ohll}[1]{\hat{\bar{\bar{{#1}}}}}

\newcommand{\oo}[1]{\overset{\circ{}}{#1}}
\newcommand{\ool}[1]{\overset{\circ{}}{\bar{{#1}}}}
\newcommand{\ooh}[1]{\overset{\circ{}}{\hat{{#1}}}}
\newcommand{\oot}[1]{\overset{\circ{}}{\tilde{{#1}}}}
\newcommand{\oohl}[1]{\overset{\circ{}}{\hat{\bar{{#1}}}}}
\newcommand{\ooht}[1]{\overset{\circ{}}{\hat{\tilde{{#1}}}}}
\newcommand{\oolt}[1]{\overset{\circ{}}{\bar{\tilde{{#1}}}}}
\newcommand{\oohlt}[1]{\overset{\circ{}}{\hat{\bar{\tilde{{#1}}}}}}

\newcommand\qtw{$q$-{\rm P}$_{\rm{II}}$ }
\newcommand\qth{$q$-{\rm P}$_{\rm{III}}$ }
\newcommand\qfo{$q$-{\rm P}$_{\rm{IV}}$ }

\newcommand{\eqn}[1]{(\ref{#1})}

\newcommand\Pa{Painlev\'e }
 \newcommand\BT{B\"acklund transformation}
 \newcommand\BP{B\"acklund parameter}

\newcommand\hIII{H3}

\newtheorem{theorem}{Theorem}
\begin{document}
\title[Geometric Reductions of ABS equations]{Geometric Reductions of ABS equations on an $n$-cube to discrete Painlev\'e systems }
\author{N. Joshi}\thanks{This research was supported by an Australian Laureate Fellowship \# FL 120100094 and grant \# DP130100967 from the Australian Research Council. }
\address{School of Mathematics and Statistics F07, The University of Sydney, NSW 2006, Australia\\ Tel: +61 2 9351 2172\\ Fax: +61 2 9351 4534}
\email{nalini.joshi@sydney.edu.au}

\author{N. Nakazono}
\address{School of Mathematics and Statistics F07, The University of Sydney, NSW 2006, Australia\\ Fax: +61 2 9351 4534}
\email{nobua.n1222@gmail.com}

\author{Y. Shi}
\address{School of Mathematics and Statistics F07, The University of Sydney, NSW 2006, Australia\\ Fax: +61 2 9351 4534}
\email{yshi7200@gmail.com}
\maketitle
\date{02 May 2014}                                      

\begin{abstract}
In this paper, we show how to relate $n$-dimensional cubes
on which ABS equations hold to the symmetry groups of discrete Painlev\'e equations. We here focus on the reduction from the 4-dimensional cube to the $q$-discrete third Painlev\'e equation, which is a dynamical system on a rational surface of type $A_5^{(1)}$ with the extended affine Weyl group $\widetilde{\mathcal W}\bigl((A_2+A_1)^{(1)}\bigr)$. We provide general theorems to show that this reduction also extends
to other discrete Painlev\'e equations at least of type A.
\end{abstract}

\maketitle

\section{Introduction}
We present a geometric method to obtain discrete Painlev\'e equations from higher-dimensional integrable discrete systems.  Geometrically, symmetry groups of discrete Painlev\'e equations are affine Weyl groups, orthogonal to the divisor class  of their initial value space in the Picard lattice \cite{oka:79,sak:01}. Higher dimensional discrete integrable systems arise from an entirely different geometric point of view, namely as multi-dimensionally consistent quad-equations embedded on a hypercube (we refer to the $n$-dimensional hypercube as the $n$-cube) \cite{nw:01, abs:03, adler2009discrete}. 

Previous studies in the literature have performed reductions of such equations via methods suited to specific examples \cite{NijP:91,Gramani:05,jos:06,Hay:07,fjn:08,O:12, HHS:13, O:14}. In particular, the identification of the reduced system
has been mainly achieved by comparing or transforming it to known forms of the discrete Painlev\'e equations. It
has been shown for systems such as the KP hierarchy and UC hierarchy that reduction from a
higher-dimensional setting is more natural \cite{KNY:2002,Tsuda:09, WH:06, Tsuda:12}. We demonstrate here that it is indeed also the case
for the higher-dimensional quad-equations consistent on the cube by providing a general geometric construction. 

Multi-dimensionally consistent equations, with copies of the same equation holding on each face of a 3-cube (a symmetric 3-cube), were classified by Adler {\it et al.} \cite{abs:03}. This is often referred to as consistency around the cube property. The results can be naturally extended to a symmetric n-cube and are called  ABS equations. Boll \cite{Boll:11} extended ABS equations to asymmetric 3-cubes, where equations on different faces may differ \cite{Atkinson:08,Boll:12}. These results were further extended to 4-dimensional cubes with the exception of asymmetric systems that incorporate  $H^6$-type equations \cite{Boll:12}.


In this work, we construct a 4-dimensionally consistent system that does incorporate $H^6$-type equations and show that the geometrical nature of the construction gives us naturally the symmetry of its periodically reduced systems, not
limiting to only reductions on 2-dimensional lattice. We show explicitly the relation of its reduced systems to the $q$-Painlev\'e equations on $A_5^{(1)}$-surface of Sakai's classification.

Our main idea comes from
the identification of the orthogonal projection of an $n$-cube in $\mathbb{R}^{n}$ with the Voronoi cell of the 
$(n-1)$-dimensional root lattice of type $A_{n-1}$. In particular, we describe the dynamics of multi-dimensionally consistent quad-equations\footnote{We use the term quad-equation throughout the paper to describe partial difference equations that relate the values of the solution on the vertices of a quadrilateral. } on the $n$-cube
in the cubic lattice, $\mathbb{Z}^n$,  by using the translations of Voronoi cells in the weight lattice of the extended affine Weyl group $\widetilde{\mathcal W}\bigl(A_{n-1}^{(1)}\bigr)$. For conciseness, we state our main results here without providing details of the proofs. Details will be given in a subsequent paper.

This work is motivated by our previous findings \cite{jns:14}, where quad-equations were observed on
what is called the $\omega$-lattice, constructed from the $\tau$-function theory of the $A_5^{(1)}$-surface $q$-Painlev\'e system. The present paper begins at the other end of the story with quad-equations on an $n$-cube.
We travel the other way to show how to construct higher-dimensional integrable systems from which 
$q$-Painlev\'e systems and extensions can be obtained along with their full parameters.

The plan of the paper is as follows. In Section 2, we state the main idea of the $n$-cube, the 
Voronoi cell of the root lattice $A_{n-1}$ and the quad-equations on an $n$-cube in Theorems 1 and 2. We
give
an explicit example for the case $n=3$. We show how to obtain the symmetry of (1,1,1)-periodically reduced
quad-equations on a
symmetric 3-cube. This information is then used in the next section as the part of construction
of a system of quad-equations on an asymmetric 4-cube.
In Section 3,
we construct an asymmetric system of consistent quad-equations on a 4-cube by fitting eight 3-cubes in a self-consistent way. In Section 4, we show how to obtain $A_5^{(1)}$-surface $q$-Painlev\'e system by imposing a (1,1,1)-periodic condition along a symmetric 3-cube 
inside of the asymmetric 4-cube. We give also the subcase of the (1,1,1)-periodic condition, namely the (2,1) periodic reduction. 
The latter example shows that our geometric approach on a higher dimensional setting 
includes periodic-type reductions on a 2D lattice approach. Finally, the paper ends with a conclusion. 
\section{The $n$-cube and the Voronoi cell of the root lattice $A_{n-1}$}

We first recall some notations and definitions needed to describe our results.
The root lattice $A_{n-1}$ is the $\mb{Z}$-span of the simple roots $\rho_i=\e_i-\e_{i+1}$, $1\leq i \leq n-1$ of the root system of type $A_{n-1}$, the corresponding Weyl group is
$\mathcal W\bigl(A_{n-1}\bigr)=\langle s_1, ..., s_{n-1}\rangle=\mathfrak S_{n}$, where $\mathfrak S_{n}$ 
denotes the
symmetric group, which acts by permuting the $\e_{i}$. The fundamental weights $h_i$, $1\leq i \leq n-1$
are defined by the inner product
$
(h_i, \rho_j)=\delta_{ij}.
$

For systems of type $A_{n-1}$, the fundamental weights are defined by
\beqn\label{hk}
h_{k}=(\e_1+...+\e_k)-\frac{k}{n}\sum_{i=1}^{n}\e_{i},\;\;1\leq k \leq n-1 .
\eeqn
The weight lattice of type $A_{n-1}$ is the $\mb {Z}$-span of the fundamental weights
$P(A_{n-1})=\mb{Z}\{h_1, ..., h_k\}$.
The Voronoi cell $V({\boldmath0})$ is the convex hull of 
\beqn \label{ws}
\{w_{S}\}=\bigcup_{0\leq k \leq |S|; w\in \mathfrak S_{n} } w (h_k),
\eeqn
where $S\subseteq\{1, ..., n\}$ and we have set $h_n=h_0=w_0={\boldmath0}$. $V(0)$ tessellates
$P(A_{n-1})$ by translations \cite{moody:92, CSbook}. The highest root of the root system of type $A_{n-1}$
 is 
\beqn\label{hr}
\ot\rho=\sum_{i=1}^{n-1}\rho_{i}=h_1+h_{n-1},
\eeqn
where $\rho_{i}$ are the simple roots. The extended affine Weyl group $\widetilde{\mathcal W}\bigl(A_{n-1}^{(1)}\bigr)$
 has generators $\langle s_0, s_1,...,  s_{n-1}, \pi\rangle$, which satisfy the following relations:
\begin{subequations}
\beqna\label{geneaW}
&&s_i^2=1,\;\;(s_is_{i+1})^3=1,\;\;(i\in\mb Z/n\mb Z)\\
&&\pi^n=1, \;\;\pi s_i=s_{i+1}\pi.
\ee 
The generators of the finite Weyl group $s_i$, $1\leq i \leq n-1$ act on $h_i$ to give the $\mathcal W$-orbit of $h_i$  and
\beqna
&&s_{i}h_k=h_k,\; j\neq k,
\ee
and 
\beqna\label{authk}
&&\pi(h_k)=h_{k+1},\;(k\in\mb Z/n\mb Z).
\ee
By definition \cite{Hbook}
\beqn
s_0(v)=s_{\ot
\rho,1}(v)=v-\left((v,\ot
\rho)-1\right)\ot
\rho,\; v\in\mb R^n,
\eeqn
therefore we have
\beqn\label{s0hk}
s_0(h_k)=h_k \;\mbox{for}\;  k\neq 0,\;\mbox{and}\; s_0(h_0)=\ot
\rho=h_1+h_{n-1}.
\eeqn

The extended affine Weyl group can be represented as the semidirect product of the finite Weyl group and the translation group corresponding to the weight lattice $\widetilde{\mathcal W}\bigl(A_{n-1}^{(1)}\bigr)=\mathcal W \bigl(A_{n-1}\bigr)\ltimes P(A_{n-1})$. Let
 $T_j$, $1\leq j \leq n$,
denotes translation in the $j$-th direction of the $n$-dimensional representation of $A_{n-1}$ in $\mathbb{R}^n$ \cite{Nbook}, 
\beqna\nonumber
&&T_1=\pi s_{n-1}...s_1,\\\nonumber
&&T_2=s_1\pi s_{n-1}...s_{2},\\\label{DefT}
&&\hspace{1mm}\vdots\\\nonumber
&&T_n=s_{n-1}...s_1\pi,
\ee
\begin{equation}\label{periodic}
T_n ... T_1=1.
\end{equation}
\end{subequations}

The $n$-cube is a combinatorial object, which can be embedded in $\mathbb{R}^n$ as follows: $x_{S}=\sum_{i\in S} \e_{i}$,
where $\e_i$ are the unit vectors of $\mathbb{R}^n$ and $S\subseteq\{1, ..., n\}$. There is a unique vertex $\xi$, which is ``furthest" from $x_{0}$, being $n$ steps away: $\xi=\sum_{i=1}^{n} x_{i}$.
The condition $\xi=x_{0}$ is equivalent to the orthogonal projection $\phi$ of the $n$-cube w.r.t $\xi$:
\beqn\label{phi}
\phi(x_S)=v-\frac{(x_S,\xi)\,\xi}{\|\xi\|}.
\eeqn

\begin{theorem}\label{t1}
The convex hull of $\phi(x_S)$, $S\subseteq\{1, ..., n\}$, is the Voronoi cell $V({\boldmath0})$  around the origin of $A_{n-1}$ root lattice. That is, $\phi(x_S)=w_S$.
 \end{theorem}

\begin{theorem}\label{t2}
The system of quad-equations
\beqn\label{H3ncube}
\frac{w_{S+\e_i}}{w_{S+\e_j}}=\frac{\al_iw_S+\al_jw_{S+\e_i+\e_j}}{\al_jw_S+\al_iw_{S+\e_i+\e_j}},
\;\;S\subset\{1, ..., n\},\;i, j \nin S,\; 0\leq |S|\leq n-2,
\eeqn
under the {\rm{(1,...,1)}} periodic condition 
\begin{equation}\label{pd}
T_n...T_1 w_S=w_S,
\end{equation}
is invariant under the symmetry group $\widetilde{\mathcal W}\bigl(A_{n-1}^{(1)}\bigr)$, where
$S+\e_i$ refers to a vertex on the $n$-cube, $\{w_{S}\}$ are the variables of the quad-equation, defined
on the vertices of $V(0)$.
The system of quad-equations {\rm \eqn{H3ncube}} are known as the $\hIII_{\delta=0}$ equation in the ABS
classification. (For brevity we denote it here by $H3$).

Furthermore, we have the following:
\begin{description}
\item[(1)] 
Let $a_j=\frac{\al_{j+1}}{\al{j}}$, $1\leq j\leq n-1$, and define $a_0=q\frac{\al_1}{\al_n}$, so that
$a_0a_1...a_{n-1}=q$,  where $q$ is a constant. The reflection generators of the finite Weyl group $s_i$ , $1\leq i \leq n-1$, act on the variables of the quad-equations $w_{S}$ by permuting the indices, 
and their actions on the parameters are defined as follows:
\begin{equation}\label{siC}
s_i(a_j)=a_ja_i^{-A_{ij}}, \;\;1\leq i, j\leq n-1,
\end{equation}
where $A_{ij}$ is the entry of the Cartan matrix of type
$A_{n-1}$:

$A=(A_{ij})_{i,j=1}^{n-1}
 =\begin{pmatrix}2&-1&&0\\-1&2&\ddots&\\&\ddots&\ddots&-1\\0&&-1&2 \end{pmatrix}.$

The actions of the
translations on the parameters and the $w_S$ variables are defined by
\begin{equation}\label{transact}
T_{j}(w_S)=w_{S+\e_j},\;\;
T_{j}(a_i)=
\begin{cases}
a_i/q,\;\;j=i\\
qa_i,\;\;j=i+1\\
a_i,\;\;j\neq i, i+1
\end{cases},
\end{equation}
where $i,j \in \mathbb{Z}/n\mathbb{Z}$. This corresponds to the tessellation of $\mathbb{R}^{n-1}$ by translations of $V({\boldmath0})$ in the weight lattice.

\item[(2)]

All of the $2^{n}$ $w_{S}$ variables can be expressed in terms of the $n$ initial values which correspond
to the fundamental weights $h_k$, $0\leq k\leq n-1$ defined in Equations {\rm{(\ref{hk}}}-{\rm{\ref{ws}})}.
The actions of the generators $\pi$ and $s_0$ can be obtained from those of the finite reflections
and translations. In particular, from Equations {\rm{(\ref{DefT}})} and {\rm{(\ref{siC}}}-{\rm{\ref{transact}})} we have
\begin{equation}
\pi=T_1s_1...s_{n-1},
\end{equation}
and
\beqna\nonumber
&&\pi(a_i)=a_{i+1},\\
&&\pi(h_k)=h_{k+1}, \;k\in \mathbb{Z}/n\mathbb{Z},
\ee
where action of $\pi$ on the initial values $h_k$ is given by Equation {\rm{(\ref{authk}})}.

From Equation {\rm{(\ref{s0hk}})}
we see that the action of $s_0$
on the initial values $h_k$, $1\leq k\leq n-1$ are trivial except on $h_0$, whose action can
be derived from Equations {\rm{(\ref{H3ncube}})} and {\rm{(\ref{transact}})}:
\begin{equation}\label{s0w0}
s_0(h_0)=h_0\frac{h_{n-1}+h_{1}/a_0}{h_{n-1}+a_0h_1}.
\end{equation}
Using the definition
\begin{equation}
s_0=\pi^{-1}s_1\pi,
\end{equation}
we have
\beqna\nonumber
&&s_0(a_{n-1})=a_{n-1}a_0,\\
&&s_0(a_0)=1/a_0,\\\nonumber
&&s_0(a_1)=a_1a_0.
\ee
\end{description}
\end{theorem}

Theorems \ref{t1} and \ref{t2} provide a general method for constructing systems of multi-dimensionally consistent quadrilateral equations and simultaneously provide the symmetry groups of its periodic reductions on an $n$-cube. We call this method the ``geometric reduction'' of such quad-equations on an $n$-cube. In the rest of this section we give
an explicit application of Theorems \ref{t1} and \ref{t2} in the case $n=3$.
\subsection{ $\widetilde{\mathcal W}\bigl(A_2^{(1)}\bigr)$ symmetry of (1,1,1) periodically reduced quad-equations on a symmetric $3$-cube}
For the case $n=3$, $x_{\{1,2,3\}}$
are the $8$ vertices of the $3$-cube: $x_0=0, x_{1}=\e_1, x_{2}=\e_2, x_{3}=\e_3, x_{12}=\e_1+\e_2, x_{13}=\e_1+\e_3, x_{23}=\e_2+\e_3, x_{123}=\e_1+\e_2+\e_3=\xi$. 

The extended affine Weyl group of type $A_2$ is
$\widetilde{\mathcal W}\bigl(A_2^{(1)}\bigr)=\langle s_0, s_1, s_2, \pi\rangle$,
with the relations
\begin{subequations}
\beqna
&&s_j^2=1,\;\;(s_js_{j+1})^3=1,\;\;(j=0, 1, 2),\\
&&\pi^3=1, \;\;\pi s_j=s_{j+1}\pi,
\ee 

and translations $T_i~(i=1,2,3)$ 
\begin{equation}
 T_1=\pi s_2s_1,\quad
 T_2=s_1\pi s_2,\quad
 T_3=s_2s_1\pi,\quad T_1T_2T_3=1.
\end{equation}
\end{subequations}
$\widetilde{\mathcal W}\bigl(A_2^{(1)}\bigr)$
 has a representation in $\mathbb{R}^{3}$ as follows:
the simple roots $\rho_1=\e_1-\e_2$, $\rho_2=\e_2-\e_3$, and the highest root is $\ot\rho=\e_1-\e_3$. The two fundamental weights $h_1=w_1$, $h_2=w_{12}$. Their respective 
$\mathcal W$-orbits are:
$w_1=\frac{1}{3}(2\e_1-\e_2-\e_3)$, $s_1(w_1)=w_2=\frac{1}{3}(-\e_1+2\e_2-\e_3)$, 
$s_2s_1(w_1)=w_3=\frac{1}{3}(-\e_1-\e_2+2\e_3)$;
$w_{12}=\frac{1}{3}(\e_1+\e_2-2\e_3)$, $s_2(w_{12})=w_{13}=\frac{1}{3}(\e_1-2\e_2+\e_3)$, 
$s_1s_2(w_{12})=w_{23}=\frac{1}{3}(-2\e_1+\e_2+\e_3)$.

The orthogonal projection $\phi$, Equation (\ref{phi}) maps the $3$-cube to the Voronoi cell of $A_2$, $\phi(x_{\{1,2,3\}})=w_{\{1,2,3\}}$:
\beqna\nonumber
&&\phi(x_1)=w_1,\;
\phi(x_2)=w_2,\;
\phi(x_3)=w_3,\\\nonumber
&&\phi(x_{12})=w_{12},\;
\phi(x_{13})=w_{13},\;
\phi(x_{23})=w_{23},
\ee
and $\phi(x_{123})=w_{123}=w_0=\phi(x_0)$.
Thus the orthogonal projection of the $3$-cube gives
the $(1, 1, 1)$ ``periodic condition'': $w_{123}=w_0$ of the quad-equations, which comes from the relation $T_1T_2T_3=1$ of $\widetilde{W}(A_2^{(1)})$.


The fundamental simplex of $A_{2}^{(1)}$ lattice is the convex hull of $w_0$,$w_1$, $w_{12}$, i.e. an equilateral  triangle $F$ bounded by the reflection hyperplanes described by the reflection generators $\langle s_0, s_1, s_2\rangle$, and
$\pi$ acts by rotating anti-clockwise $120^\circ{}$ around the barycenter of $F$. We obtain the 
Voronoi cell at the origin $V(0)$ by applying reflections of $\mathcal W\bigl(A_2\bigr)$ to $F$, and
we can cover the whole $2$-D plane by translating $V(0)$ \cite{Nbook}.
$\mathcal W\bigl(A_{2}\bigr)$ is the symmetric group $\mathfrak S_3=\{1, s_1, s_2, s_1s_2, s_2s_1, s_1s_2s_1=s_2s_1s_2\}$,  that is $|\mathfrak S_3|=6$.

\begin{figure}[ht!] 
\centering
  \begin{tikzpicture}[scale=2]
 		\foreach \s in {2,3,4,5}
		{
\draw[dashed] (0+\s,0) -- (0.5+\s, 0.866025403784);
\draw[dashed] (1+\s, 0.866025403784) -- (-1+\s, 0.866025403784);
\draw[dashed] (-0.5+\s, 0.866025403784) -- (-1+\s,0) ;
\draw[dashed] (-1+\s,0) -- (\s,0);
\draw[dashed] (-0.5+\s, -0.866025403784) -- (0.5+\s, -0.866025403784) ;
\draw[dashed] (-0.5+\s, 0.866025403784)--(0+\s, 0);

               }

\foreach \s in {2,3,4}
		{
\draw[dashed] (\s,0) -- (0.5+\s, -0.866025403784);
\draw[dashed] (\s,0) -- (-0.5+\s, -0.866025403784);
\draw[dashed] (0.5+\s, -0.866025403784) -- (1+\s,0);
\draw[dashed] (1+\s,1.73205080757) -- (0.25+\s, 2.89807621135);
\draw[dashed] (0+\s,1.73205080757) -- (0.5+\s, 2.59807621135);
\draw[dashed] (0.5+\s,2.59807621135) -- (0.75+\s, 3.03105);
\draw[dashed] (\s,1.73205080757) -- (-0.5+\s, 0.866025403784);
		}

 		\foreach \s in {2,3,4,5}
		{

\draw[dashed] (0.5+\s, 2.55807621135) -- (-0.5+\s, 2.55807621135);
\draw[dashed] (-1+\s,1.73205080757) -- (\s,1.73205080757);
\draw[dashed] (-0.5+\s, 0.866025403784) -- (0.5+\s, 0.866025403784) ;
\draw[dashed] (\s,1.73205080757) -- (0.5+\s, 0.866025403784);
\draw[dashed] (0.5+\s, 0.866025403784) -- (1+\s,1.73205080757);
		}
\draw[thick] (3, 0) -- (2.5, 0.866025403784) ; 
\draw[thick] (2.5, 0.866025403784) -- (3, 1.73205080757)  ;
\draw[thick] (3, 1.73205080757) -- (4, 1.73205080757)  ;
\draw[thick] (4, 1.73205080757) -- (4.5, 0.866025403784);
\draw[thick] (4.5, 0.866025403784) -- (4, 0)  ;
\draw[thick] (4, 0) -- (3, 0)  ;

\filldraw [black]  (6.5, 0.866025403784) node   {\colorbox{white}{$s_2$}} circle(0.1ex);
\filldraw [black]  (4.75, 3.13205080757) node  {\colorbox{white}{$s_1$}} circle(0.1ex);
\filldraw [black]  (3.2, 3.13205080757) node  {\colorbox{white}{$s_0$}} circle(0.1ex);

\filldraw [black]  (4, 0.53) node  {\colorbox{white}{{\footnotesize$s_2F$}}} circle(0.1ex);
\filldraw [black]  (4, 1.23) node  {\colorbox{white}{{\footnotesize$F$}}} circle(0.1ex);
\filldraw [black]  (3.5, 0.33) node  {\colorbox{white}{{\footnotesize$s_2s_1F$}}} circle(0.1ex);
\filldraw [black]  (3.5, 1.53) node  {\colorbox{white}{\footnotesize{$s_1F$}}} circle(0.1ex);
\filldraw [black]  (3, 0.63) node  {\colorbox{white}{{\footnotesize$s_2s_1s_2F$}}} circle(0.1ex);
\filldraw [black]  (3, 1.03) node  {\colorbox{white}{{\footnotesize$s_1s_2F$}}} circle(0.1ex);
\filldraw [black]  (5, 1.23) node  {\colorbox{white}{{\footnotesize$T_1F$}}} circle(0.1ex);

\filldraw [black]  (3.5, 0.866025403784) node  {\colorbox{white}{$w_0$}} circle(0.1ex);
\filldraw [black]  (4.5, 0.866025403784) node  {\colorbox{white}{$w_1$}} circle(0.1ex);
\filldraw [black]  (3, 1.73205080757) node  {\colorbox{white}{$w_2$}} circle(0.1ex);
\filldraw [black]  (3, 0) node  {\colorbox{white}{$w_3$}} circle(0.1ex);
\filldraw [black]  (2.5, 0.866025403784) node  {\colorbox{white}{$w_{23}$}} circle(0.1ex);
\filldraw [black]  (4, 1.73205080757) node  {\colorbox{white}{$w_{12}$}} circle(0.1ex);
\filldraw [black]  (4, 0) node  {\colorbox{white}{$w_{13}$}} circle(0.1ex);
\filldraw [black]  (5, 1.73205080757) node  {\colorbox{white}{$w_{112}$}} circle(0.1ex);
\filldraw [black]  (5.5, 0.866025403784) node  {\colorbox{white}{$w_{11}$}} circle(0.1ex);

\draw (5.9, 0.866025403784) node [rotate=-180] {$\curvearrowupdown$};
\draw (3.27, 2.83205080757) node [rotate=-45] {$\curvearrowupdown$};
\draw (4.66, 2.83205080757) node [rotate=-135] {$\curvearrowupdown$};

\end{tikzpicture}
\caption{ Voronoi cell of $A_2^{(1)}$ lattice, a hexagon.}\label{a2V lattice}
\end{figure}
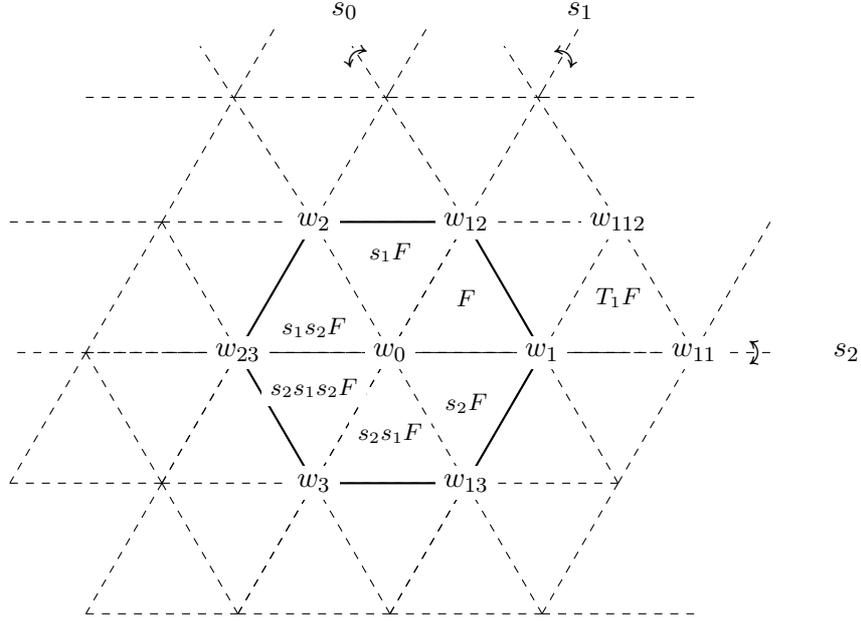

From 
\beqn
V({\boldmath0})=\bigcup_{w\in \mathfrak S_3} w F,
\eeqn
we see that $V({\boldmath0})$ is a hexagon made up of the union of six equilateral triangles. See Figure \ref{a2V lattice}.

The quad-equations on the $3$-cube are the realizations of the reflections in $\mathcal W\bigl(A_2^{(1)}\bigr)$:
\beqna
&&s_0(w_0)=w_{112}=\frac{w_{0}(q \al w_1+\ga w_{12})}{\ga w_1+q \al w_{12}}\\
&&s_1(w_1)=w_{2}=\frac{w_{1}(\al w_0+\be w_{12})}{\be w_0+\al w_{12}}\\
&&s_2(w_{12})=w_{13}=\frac{w_{12}(\be w_1+\ga w_{0})}{\ga w_1+\al w_{0}},
\ee
where $w_1$, $w_{12}$ and $w_0$  are the 3 initial values of the system of quad-equations, corresponding to
the fundamental weights $h_1$, $h_2$ and $h_0$.
Define 
\begin{equation}\label{abg}
a_1=\frac{\be}{\al},\;
a_2=\frac{\gamma}{\be},\;
a_0=\frac{q\al}{\gamma},
\end{equation} we have
\begin{equation}\label{sa}
s_i(a_j)=a_ja_i^{-A_{ij}}, \;\;0\leq i, j\leq 2,
\end{equation}
where $A_{ij}$ is the entry of the Cartan matrix of type
$A_{2}^{(1)}$:

$A=(A_{ij})_{i,j=0}^{2}
 =\begin{pmatrix}2&-1&-1\\-1&2&-1\\-1&-1&2 \end{pmatrix}.$

The actions of $\pi$ are:
\beqna
&&\pi(w_0)=w_1,\; \pi(w_1)=w_{12},\; \pi(w_{12})=w_0,\\
&&\pi(a_0)=a_1,\; \pi(a_1)=a_{2},\; \pi(a_{2})=a_0.
\ee

Translations of $\widetilde{W}(A_2^{(1)})$ act on the
vertices of the Voronoi cell by translating them in the directions along $w_1$, $w_2$
and $w_3$. For example,

\beqna\nonumber
&&T_1(w_0)=w_1,\\
&&T_1(w_1)=w_{11},\\\nonumber
&&T_1(w_{12})=w_{112}.
\ee
 and actions on the parameters are

\beqna\nonumber
&&T_1(a_1)=a_1/q,\; T_2(a_1)=qa_1,\\
&&T_2(a_2)=a_2/q,\; T_3(a_2)=qa_2,\\\nonumber
&&T_3(a_0)=a_0/q,\; T_1(a_0)=qa_0.
\ee

We here see that the (1,1,1) reduction of the symmetric 3-cube with $\hIII$ equations
results in a system with $\widetilde{\mathcal W}\bigl(A_2^{(1)}\bigr)$ symmetry. It is known
that the full symmetry of the $q$-discrete \Pa equation associated with surface type $A_5^{(1)}$ has $\widetilde{\mathcal W}\bigl((A_2+A_1)^{(1)}\bigr)$ symmetry. In the next section we construct 
the higher dimensional system, in particular, quad-equations consistent on an asymmetric 4-cube. By reduction, this will give rise to exactly the $q$-discrete \Pa equation
with $\widetilde{\mathcal W}\bigl((A_2+A_1)^{(1)}\bigr)$ symmetry.
\section{Equations on an asymmetric 4-cube}
To construct a 4-cube, we need four lattice directions $l, m, n, k$ and four lattice parameters $\al(l), \be(m), \ga(n), \la(k)$.
Let the dependent variable be $x=x(l, m, n, k)$ and denote its shifts in each direction respectively by
\begin{subequations}
\begin{align}
&\ol x=x_1=x(l+1,m,n,k),\\
&\oh x=x_2=x(l,m+1,n,k),\\
&\ot x=x_3=x(l,m,n+1,k),\\
&\oo x=x_4=x(l,m,n,k+1).
\end{align}
\end{subequations}
\subsection{Symmetric 3-cube $\mathbf C_{312}$}
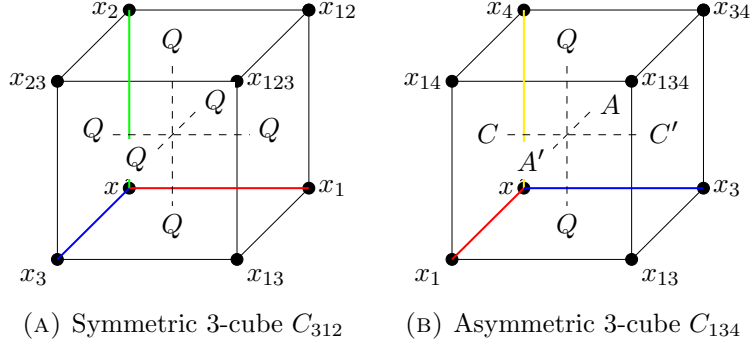
\begin{figure}[ht!]
\begin{subfigure}[b]{0.40\textwidth}
\centering
        \resizebox{\linewidth}{!}{
\begin{tikzpicture}[scale=0.5]	
\filldraw [black] (0,0) node [anchor=north east] {$x_3$}  circle(1ex);
\filldraw [black]  (0,5) node [anchor=east] {$x_{23}$ } circle(1ex);
\filldraw [black] (5,0) node [anchor=north west] {$x_{13}$ } circle(1ex);
\filldraw [black](5,5) node [anchor=west] {$x_{123}$ } circle(1ex);

\filldraw [black]  (2,2) node [anchor=east] {$x$} circle(1ex);
\filldraw [black] (2,7) node [anchor=east] {$x_2$} circle(1ex);
\filldraw [black](7,2) node [anchor=west] {$x_1$} circle(1ex);
\filldraw [black](7,7) node [anchor=west] {$x_{12}$ } circle(1ex);
\filldraw [black](7,7)  circle(1ex);

\draw (0,0) -- (0,5);
\draw (0,0) -- (5,0);
\draw (5,0) -- (5,5);
\draw (0,5) -- (5,5);

\draw (5,0) -- (7,2);
\draw (5,5) -- (7,7);
\draw (0,5) -- (2,7);
\draw (2,7) -- (7,7);
\draw (7,2) -- (7,7);

\draw[thick,blue] (0,0) -- (2,2);
\draw[thick,green] (2,2) -- (2,7);
\draw[thick,red] (2,2) -- (7,2);
\draw[dashed] (3.2, 1) -- (3.2, 6);
\draw (3.2, 5.3) node [anchor=south] {\colorbox{white}{$Q$}};
\draw (3.2,1.75) node [anchor=north] {\colorbox{white}{$Q$}};
\draw[dashed] (0.7, 3.5) -- (5.7, 3.5);
\draw (5.9, 2.8) node [anchor=south] {\colorbox{white}{$Q$}};
\draw (1,2.8) node [anchor=south] {\colorbox{white}{$Q$}};
\draw[dashed] (1.9, 2.2) -- (4.17642353761, 4.47642353761);
\draw (4.4, 3.6) node [anchor=south] {\colorbox{white}{$Q$}};
\draw (2.2,2) node [anchor=south] {\colorbox{white}{$Q$}};
\end{tikzpicture}
}
\subcaption{Symmetric 3-cube $C_{312}$}\label{fC312}
\end{subfigure}
\begin{subfigure}[b]{0.40\textwidth}
\centering
        \resizebox{\linewidth}{!}{
\begin{tikzpicture}[scale=0.5]	
\filldraw [black] (0,0) node [anchor=north east] {$x_1$}  circle(1ex);
\filldraw [black]  (0,5) node [anchor=east] {$x_{14}$ } circle(1ex);
\filldraw [black] (5,0) node [anchor=north west] {$x_{13}$ } circle(1ex);
\filldraw [black](5,5) node [anchor=west] { $x_{134}$} circle(1ex);

\filldraw [black]  (2,2) node [anchor=east] {$x$} circle(1ex);
\filldraw [black] (2,7) node [anchor=east] {$x_4$} circle(1ex);
\filldraw [black](7,2) node [anchor=west] {$x_3$} circle(1ex);
\filldraw [black](7,7) node [anchor=west] {$x_{34}$ } circle(1ex);
\filldraw [black](7,7)  circle(1ex);

\draw (0,0) -- (0,5);
\draw (0,0) -- (5,0);
\draw (5,0) -- (5,5);
\draw (0,5) -- (5,5);

\draw (5,0) -- (7,2);
\draw (5,5) -- (7,7);
\draw (0,5) -- (2,7);
\draw (2,7) -- (7,7);
\draw (7,2) -- (7,7);

\draw[thick,red] (0,0) -- (2,2);
\draw[thick,yellow] (2,2) -- (2,7);
\draw[thick,blue] (2,2) -- (7,2);
\draw[dashed] (3.2, 1) -- (3.2, 6);
\draw (3.2, 5.3) node [anchor=south] {\colorbox{white}{$Q$}};
\draw (3.2,1.75) node [anchor=north] {\colorbox{white}{$Q$}};
\draw[dashed] (0.7, 3.5) -- (5.7, 3.5);
\draw (5.9, 2.8) node [anchor=south] {\colorbox{white}{$C'$}};
\draw (1,2.8) node [anchor=south] {\colorbox{white}{$C$}};
\draw[dashed] (1.9, 2.2) -- (4.17642353761, 4.47642353761);
\draw (4.4, 3.6) node [anchor=south] {\colorbox{white}{$A$}};
\draw (2.2,2) node [anchor=south] {\colorbox{white}{$A'$}};

\end{tikzpicture}
}
\caption{Asymmetric 3-cube $C_{134}$\label{fC134}}
\end{subfigure}
\caption{Two types of 3-cubes $C_{312}$ and $C_{134}$ with different arrangements of equations on faces}.
\label{two 3cubes}
\end{figure}

We start with the $\hIII$  quad-equation \eqref{f13s} 
on a face of the 4-cube given by the $l$, $n$ directions,
where $x=x(l,n)$, $x_1=x(l+1,n)$, $x_3=x(l,n+1)$, $x_{13}=x(l+1,n+1)$. 
We construct a symmetric system of equations \eqn{f13s}--\eqn{f132} on a 3-cube by taking auto-\BT s\ of equation \eqref{f13s}: 
\begin{subequations}\label{f13}
\begin{align}\label{f13s}
&Q(x,x_{1},x_{3},x_{13};\al, \ga)=xx_3+x_1x_{13}-\al/\ga(xx_1+x_3x_{13})=0,\\\label{f12}
&Q(x,x_1, x_2, x_{12} ;\al,\be)=0,\\\label{f23}
&Q(x,x_3, x_2, x_{23} ;\ga,\be)=0,\\\label{f123}
&Q(x_1, x_{13}, x_{12}, x_{123} ;\ga,\be)=0,\\\label{f132}
&Q(x_2,x_{12}, x_{23}, x_{123} ;\al,\ga)=0,
\end{align}
\end{subequations}
where $x_2$ denotes the auto-\BT \ of $x$ with \BP \;$\be$.
We denote this 3-cube by $C_{312}$. See Figure \ref{fC312}.

In this system we have four initial values $x$, $x_1$, $x_2$, $x_3$;
and three quad-equations \eqn{f13s}--\eqn{f23}, adjacent to the vertex $x$.
Vertices $x_{12},\;x_{13},\;x_{23}$ are evaluated using the initial conditions
and equations \eqref{f13s}--\eqref{f23}. $x_{123}$ can then be evaluated
using equations \eqref{f123} and \eqref{f132}. 3D consistency of the system 
means that $x_{123}$
evaluated by these three equations coincide. This is true in $n$-dimensions, i.e., it is multi-dimensionally consistent and hence it
can be embedded on an $n$-cube. It is this remarkable fact we are going to
utilize in deriving discrete \Pa equations as periodic reductions on an $n$-cube. However, note that multi-dimensional consistent systems
are not limited to systems with the same equation on all the faces. 
\subsection{Asymmetric 3-cube $\mathbf C_{134}$}
Now consider a non-auto \BT\  of \eqn{f13a} giving rise to different equations on different pairs of faces on the 3-cube. Here, we use Equations from Boll's classification of asymmetric 3-cubes \cite{Boll:11c}  ((3.29)--(3.30) with $\de_2=\de_3=0$). The 3-cube contains two $H^{4}$ type ($\hIII$) equations on the ``Q'' faces and four $H^{6}$ type equations for  the``A'' and ``C'' faces.
\begin{subequations}
\begin{align}\label{f13a}
&Q(x,x_{1},x_{3},x_{13};\al, \ga):=xx_3+x_1x_{13}-\,\al/\ga(xx_1+x_3x_{13})=0,\\\label{f34}
&A(x,x_3,x_4,x_{34};\de_1)=xx_4+x_3x_{34}+\de_1xx_3=0,\\\label{f14}
&C(x,x_1,x_4,x_{14};\de_1\eta_1)=xx_4+x_1x_{14}+\de_1\eta_1xx_1=0,\\\label{f341}
&A'=A(x_1,x_{13},x_{14},x_{134};\de_1)=x_1x_{14}+x_{13}x_{134}+\de_1x_1x_{13}=0,\\\label{f143}
&C'=C(x_3,x_{13},x_{34},x_{134};\de_1\eta_1)=x_3x_{34}+x_{13}x_{134}+\de_{1}\eta_1x_3x_{13}=0,\\\label{f134}
&Q(x_4,x_{14},x_{34},x_{134};\al,\ga)=0, 
\end{align}
\end{subequations}
where $x_4$ denotes the non-auto \BT\ of $x$ and $\eta_1=\al/\ga$.  The parameter
$\de_1$ initially is a function of $l$, $m$, $n$ and $k$, to be specified 
by the consistency conditions as we construct the 4-cube. Equations \eqn{f13a}--\eqn{f134} make up the six quad-equations consistent on an asymmetric 3-cube.
We denote this 3-cube by $C_{134}$, see Figure \ref{fC134}.

Note that in obtaining Equations \eqn{f341}--\eqn{f134} we have used
\begin{equation}\label{d1l}
 \ol \de_1=\de_1,\quad
 \ot \de_{1}\frac{\al}{\ot \ga}=\de_{1}\frac{\al}{\ga}
 \ \Rightarrow \ 
 \frac{\ot \de_{1}}{\de_{1}}=\frac{\ot \ga}{\ga}.
\end{equation}

Each of the above constructions can be repeated to obtain two more 3-cubes labelled $C_{214}$ and $C_{324}$. $C_{214}$ is obtained in the same way as $C_{134}$
by replacing subscript $3$ with $2$ everywhere, with the condition $\eta_2={\ga}/{\be}$. $C_{324}$ is  obtained from $C_{214}$ by replacing subscript $1$ with $3$ everywhere, with the condition $\eta_3={\al}/{\be}$. In obtaining these two 3-cubes, we have imposed the following conditions
\begin{equation}\label{d123}
 \de_2=\de_3=\de_1\frac{\be}{\ga},\quad
 \oh \de_1=\de_1.
\end{equation}
Equations \eqn{d1l} and \eqn{d123} imply that
\beqn\label{d1k}
\de_1(n,k)=\ga(n) K(k),
\eeqn where $K$ is an arbitrary function of $k$.
\par
\begin{figure}[ht!] 
\begin{subfigure}[t]{0.40\textwidth}
\centering
        \resizebox{\linewidth}{!}{
\begin{tikzpicture}[scale=0.7]	
 	\foreach \s in {1}
	{
\draw (0+\s,\s+0.3) -- (0+\s,\s+5.3);
\draw (0+\s,\s+0.3) -- (6+\s,\s+0.3);
\draw (6+\s,\s+0.3) -- (6+\s,\s+5.3);
\draw (0+\s,\s+5.3) -- (6+\s,\s+5.3);

\draw (6+\s,\s+0.3) -- (8+\s,\s+1);
\draw (6+\s,\s+5.3) -- (8+\s,\s+6);
\draw (0+\s,\s+5.3) -- (2+\s,\s+6);
\draw (2+\s,\s+6) -- (8+\s,\s+6);
\draw (8+\s,\s+1) -- (8+\s,\s+6);

\draw (0+\s,\s+0.3) -- (2+\s,\s+1);
\draw (2+\s,\s+1) -- (2+\s,\s+6);
\draw (2+\s,\s+1) -- (8+\s,\s+1);
}
\foreach \r in {0.4}
 	{\foreach \s in {0.8}
	{
	\foreach \t in {0.6}
	{
	\draw (2+\s,2+\r) -- (2+\s,4.5+\r);
	\draw (2+\s,2+\r) -- (4.5+\s,2+\r);
	\draw (4.5+\s,2+\r) -- (4.5+\s,4.5+\r);
	\draw (2+\s,4.5+\r) -- (4.5+\s,4.5+\r);

	\draw (4.5+\s,2+\r) -- (5.5+\s,-\t+3+\r);
	\draw (4.5+\s,4.5+\r) -- (5.5+\s,-\t+5.5+\r);
	\draw (2+\s,4.5+\r) -- (3+\s,-\t+5.5+\r);
	\draw (3+\s,-\t+5.5+\r) -- (5.5+\s,-\t+5.5+\r);
	\draw (5.5+\s,-\t+3+\r) -- (5.5+\s,-\t+5.5+\r);

	\draw[very thick,blue] (2+\s,2+\r) -- (3+\s,-\t+3+\r);
	\draw[very thick,green] (3+\s,-\t+3+\r) -- (3+\s,-\t+5.5+\r);
	\draw[very thick,red] (3+\s,-\t+3+\r) -- (5.5+\s,-\t+3+\r);
	\filldraw [black]  (3+\s,-\t+3+\r) node [anchor=east] {$x$} circle(0.5ex);
	\filldraw [black]  (2+\s,2+\r) node [anchor=east] {$x_3$} circle(0.5ex);
	\filldraw [black]  (3+\s,-\t+5.5+\r) node [anchor=east] {$x_2$} circle(0.5ex);
	\filldraw [black]  (5.5+\s,-\t+3+\r) node [anchor=west] {$x_1$} circle(0.5ex);
	
	\filldraw [black]  (2+\s,4.5+\r) node [anchor=east] {$x_{23}$} circle(0.5ex);
	\filldraw [black]  (5.5+\s,-\t+5.5+\r) node [anchor=west] {$x_{12}$} circle(0.5ex);
	\filldraw [black]  (4.5+\s,2+\r) node [anchor=west] {$x_{13}$} circle(0.5ex);
	\filldraw [black]  (4.5+\s,4.5+\r) node [anchor=west] {$x_{123}$} circle(0.5ex);
	
	}
	}
}
\draw[dashed] (2+0.8,2+0.4) -- (0+1,1+0.3);
\draw[dashed] (2+0.8,4.5+0.4) -- (0+1,1+5.3);
\draw[dashed] (3+0.8,-0.6+5.5+0.4) -- (2+1,1+6);
\draw[very thick,yellow,dashed] (3+0.8,-0.6+3+0.4) -- (2+1,1+1);
\draw[dashed] (5.5+0.8,-0.6+3+0.4) -- (8+1,1+1);
\draw[dashed] (4.5+0.8,2+0.4) -- (6+1,1+0.3);
\draw[dashed] (4.5+0.8,4.5+0.4) -- (6+1,1+5.3);
\draw[dashed] (5.5+0.8,-0.6+5.5+0.4) -- (8+1,1+6);

        \filldraw [black]  (0+1,1+0.3) node [anchor=east] {$x_{34}$} circle(0.5ex);
	\filldraw [black]  (6+1,1+0.3) node [anchor=west] {$x_{134}$} circle(0.5ex);
	\filldraw [black]  (0+1,1+5.3) node [anchor=east] {$x_{234}$} circle(0.5ex);
	\filldraw [black]  (2+1,1+1) node [anchor=east] {$x_4$} circle(0.5ex);
	\filldraw [black]  (8+1,1+6) node [anchor=west] {$x_{124}$} circle(0.5ex);
	\filldraw [black]  (8+1,1+1) node [anchor=west] {$x_{14}$} circle(0.5ex);
	\filldraw [black]  (6+1,1+5.3) node [anchor=west] {$x_{1234}$} circle(0.5ex);
	\filldraw [black]  (2+1,1+6) node [anchor=east] {$x_{24}$} circle(0.5ex);
\end{tikzpicture}
}
\caption{A 4-cube.}\label{4cube}
\end{subfigure}
\begin{subfigure}[t]{0.40\textwidth}
\centering
        \resizebox{\linewidth}{!}{
\begin{tikzpicture}[scale=0.7]	
 	\foreach \s in {1}
	{
\draw (0+\s,\s+0.3) -- (0+\s,\s+5.3);
\draw (0+\s,\s+0.3) -- (6+\s,\s+0.3);
\draw (6+\s,\s+0.3) -- (6+\s,\s+5.3);
\draw (0+\s,\s+5.3) -- (6+\s,\s+5.3);

\draw (6+\s,\s+0.3) -- (8+\s,\s+1);
\draw (6+\s,\s+5.3) -- (8+\s,\s+6);
\draw (0+\s,\s+5.3) -- (2+\s,\s+6);
\draw (2+\s,\s+6) -- (8+\s,\s+6);
\draw (8+\s,\s+1) -- (8+\s,\s+6);

\draw (0+\s,\s+0.3) -- (2+\s,\s+1);
\draw (2+\s,\s+1) -- (2+\s,\s+6);
\draw (2+\s,\s+1) -- (8+\s,\s+1);
}
\foreach \r in {0.4}
 	{\foreach \s in {0.8}
	{
	\foreach \t in {0.6}
	{
\draw	
	[preaction={pattern=crosshatch}]
	[fill=pink, opacity=0.6] (3+\s,-\t+3+\r) -- (5.5+\s,-\t+3+\r)-- (5.5+\s,-\t+5.5+\r) -- (3+\s,-\t+5.5+\r)--(3+\s,-\t+3+\r);
\draw[fill=pink, opacity=0.6] (3+\s,-\t+3+\r) -- (5.5+\s,-\t+3+\r)--(8+1,1+1)--(2+1,1+1);
\draw[fill=pink, opacity=0.6] (3+\s,-\t+5.5+\r)-- (3+\s,-\t+3+\r) --(2+1,1+1)--(2+1,1+6);
\draw[fill=pink, opacity=0.6] (3+\s,-\t+5.5+\r)--(2+1,1+6)--(8+1,1+6)-- (5.5+\s,-\t+5.5+\r);
\draw[fill=pink, opacity=0.6]  (8+1,1+6)--(5.5+\s,-\t+5.5+\r)--(5.5+\s,-\t+3+\r)--(8+1,1+1);

\draw[fill=blue, opacity=0.6] (5.5+\s,-\t+5.5+\r)--(8+1,1+6)--(6+1,1+5.3)--(4.5+\s,4.5+\r);
\filldraw [black]  (3+\s,-\t+3+\r) node [anchor=east] {$x$} circle(0.5ex);
	\filldraw [black]  (2+\s,2+\r) node [anchor=east] {$x_3$} circle(0.5ex);
	\filldraw [red]  (3+\s,-\t+5.5+\r) node [anchor=east] {$x_2$} circle(0.5ex);
	\filldraw [red]  (5.5+\s,-\t+3+\r) node [anchor=west] {$x_1$} circle(0.5ex);
	
	\filldraw [black]  (2+\s,4.5+\r) node [anchor=east] {$x_{23}$} circle(0.5ex);
	\filldraw [black]  (5.5+\s,-\t+5.5+\r) node [anchor=west] {$x_{12}$} circle(0.5ex);
	\filldraw [black]  (4.5+\s,2+\r) node [anchor=west] {$x_{13}$} circle(0.5ex);
	\filldraw [black]  (4.5+\s,4.5+\r) node [anchor=west] {$x_{123}$} circle(0.5ex);

\draw (2+\s,2+\r) -- (2+\s,4.5+\r);
	\draw (2+\s,2+\r) -- (4.5+\s,2+\r);
	\draw (4.5+\s,2+\r) -- (4.5+\s,4.5+\r);
	\draw (2+\s,4.5+\r) -- (4.5+\s,4.5+\r);

	\draw (4.5+\s,2+\r) -- (5.5+\s,-\t+3+\r);
	\draw (4.5+\s,4.5+\r) -- (5.5+\s,-\t+5.5+\r);
	\draw (2+\s,4.5+\r) -- (3+\s,-\t+5.5+\r);
	\draw (3+\s,-\t+5.5+\r) -- (5.5+\s,-\t+5.5+\r);
	\draw (5.5+\s,-\t+3+\r) -- (5.5+\s,-\t+5.5+\r);

	\draw[very thick,blue] (2+\s,2+\r) -- (3+\s,-\t+3+\r);
	\draw[very thick,green] (3+\s,-\t+3+\r) -- (3+\s,-\t+5.5+\r);
	\draw[very thick,red] (3+\s,-\t+3+\r) -- (5.5+\s,-\t+3+\r);

	}
	}
}
\draw[dashed] (2+0.8,2+0.4) -- (0+1,1+0.3);
\draw[dashed] (2+0.8,4.5+0.4) -- (0+1,1+5.3);
\draw[dashed] (3+0.8,-0.6+5.5+0.4) -- (2+1,1+6);
\draw[very thick,yellow,dashed] (3+0.8,-0.6+3+0.4) -- (2+1,1+1);
\draw[dashed] (5.5+0.8,-0.6+3+0.4) -- (8+1,1+1);
\draw[dashed] (4.5+0.8,2+0.4) -- (6+1,1+0.3);
\draw[dashed] (4.5+0.8,4.5+0.4) -- (6+1,1+5.3);
\draw[dashed] (5.5+0.8,-0.6+5.5+0.4) -- (8+1,1+6);

        \filldraw [black]  (0+1,1+0.3) node [anchor=east] {$x_{34}$} circle(0.5ex);
	\filldraw [black]  (6+1,1+0.3) node [anchor=west] {$x_{134}$} circle(0.5ex);
	\filldraw [black]  (0+1,1+5.3) node [anchor=east] {$x_{234}$} circle(0.5ex);
	\filldraw [red]  (2+1,1+1) node [anchor=east] {$x_4$} circle(0.5ex);
	\filldraw [red]  (8+1,1+6) node [anchor=west] {$x_{124}$} circle(0.5ex);
	\filldraw [black]  (8+1,1+1) node [anchor=west] {$x_{14}$} circle(0.5ex);
	\filldraw [black]  (6+1,1+5.3) node [anchor=south] {$x_{1234}$} circle(0.5ex);
	\filldraw [black]  (2+1,1+6) node [anchor=east] {$x_{24}$} circle(0.5ex);
\end{tikzpicture}
}
\caption{The 3-cube $C_{214}$ is shaded pink and its tetrahedron equation is on the red vertices. Equation \eqn{f3421}
holds on the face shaded blue.}\label{red4cube}
\end{subfigure}
\caption{The colours red, blue, green and yellow denote edges that are labelled by the same respective colours in Figure \ref{two 3cubes}.}\label{two 4cubes}
\end{figure}
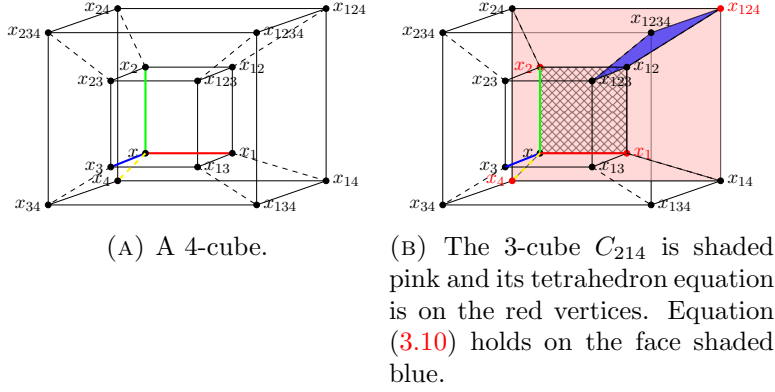

We extend the above construction to four dimensions in the following way. The cube $C_{3124}$ is obtained by shifting $C_{312}$ in the $k$ direction, $C_{1342}$ by shifting $C_{134}$ in the $m$ direction,  $C_{2143}$ by shifting $C_{214}$ in the $n$ direction and $C_{3241}$ by shifting $C_{324}$ in the $l$ direction. In summary, there are four 3-cubes adjacent to $x(l,m,n,k)$, namely, $C_{312}$, $C_{134}$, $C_{214}$ and $C_{324}$ and four 3-cubes adjacent to $x_{1234}$, namely $C_{3124}$, $C_{1342}$, $C_{2143}$ and $C_{3241}$. These four 3-cubes are all 3D consistent under the conditions \eqn{d1l} and \eqn{d123}. Hence, we have eight three dimensionally consistent 3-cubes fitted consistently in a 4-cube. See Figure \ref{4cube}. 

The result contains two \BT s, i.e., Equations \eqref{f132} and \eqref{f134}, of the equation on the bottom face of the 4-cube, namely Equation \eqref{f13s}. The permutability of these two \BT s provides a system of 24 quad-equations, which can be embedded consistently on the 24 faces of the 4-cube. By construction, the system of equations on this 4-cube is four dimensionally consistent \cite{Boll:12}.

\subsection{A (1,1,1)-reduction on the asymmetric 4-cube}
We apply a periodic condition on the asymmetric 4-cube
\beqn\label{pc}
 \ohlt x=-i\la x,  
\eeqn 
where $\la$ is only a function of $k$ and $\oo \la=q\la$. We call this a (1,1,1)-reduction of the 4-cube.
We have the following conditions on the lattice parameters in the $l$, $m$, $n$ directions: ${\ol \al}/{\al}={\oh \be}/{\be}={\ot \ga}/{\ga}=q$,
which are solved by $\alpha(l)=e^{\alpha_1}q^{l}$, $\beta(m)=e^{\beta_1}q^{m}$, $\ga(n)=e^{\ga_1}q^{n}$. 

The periodic condition \eqn{pc} enables us to evaluate $x(l,m,n,k)$ at any point in the lattice defined by 
the three directions $l$, $m$, $n$, using the initial conditions $x$, $x_1$, $x_2$, $x_3$ and the quad
equations \eqn{f13s}--\eqn{f23}.
To show that the reduced 4-cube can be extended also in the $k$ direction, i.e., we can embed the reduced 4-cube
in a four dimensional lattice, we
derive the expression relating iteration of $x(l,m,n,k)$ in the $k$ direction (i.e., $x_4$) with iterations in
two of the
$l, m, n$ directions, (for example $x_1$ and $x_2$).

On the asymmetric 3-cube $C_{214}$, shaded pink in Figure \eqref{red4cube}, we have six equations for its six faces:
\begin{subequations}
\begin{align}\label{f12a}
 &Q(x,x_1, x_2, x_{12} ;\al,\be)=0,\\\label{f24}
 &A(x,x_2,x_4,x_{24};\de_3)=xx_4+x_2x_{24}+\de_1 \be \ga^{-1} xx_2=0,\\\label{f14a}
 &C(x,x_1,x_4,x_{14};\de_3\eta_3)=xx_4+x_1x_{14}+\de_1 \al \ga^{-1}xx_1=0,\\\label{f241}
 &A'=A(x_1,x_{12},x_{14},x_{124};\de_3)=x_1x_{14}+x_{12}x_{241}+\de_1 \be \ga^{-1}x_1x_{12}=0,\\\label{f142}
 &C'=C(x_2,x_{12},x_{24},x_{124};\de_3\eta_3)=x_2x_{24}+x_{12}x_{124}+\de_1\al \ga^{-1} x_2x_{12}=0,\\\label{f124}
 &Q(x_4,x_{14},x_{24},x_{124};\al,\be)=0.
\end{align}
\end{subequations}
From \eqn{f12}, \eqn{f24} and \eqn{f142}, we find \beqn\label{C214k}
\frac{x_{12}}{x}\,\frac{x_{24}}{x_{12}}\,\frac{x}{x_{24}}=1\quad\Rightarrow\ \frac{(\al x_1-\be x_2)( x_{124}+\al K x_2)}
{(\be x_1-\al x_2)( x_4+\be K x_2)}=1.
\eeqn 
(This follows from the tetrahedron property \cite{abs:03} and by ``summing'' the three-leg forms of the equations on the faces adjacent to $x_2$.) 
Moreover, from the face shaded blue in Figure \ref{red4cube} on which holds the equation for $x_{12}$, $x_{124}$, $x_{1234}$, and $x_{123}$
 we have another expression for $x_{124}$:
\begin{equation}\label{f3421}
 x_{1234}=-\frac{x_{12}(x_{124}+\ga Kx_{123})}{x_{123}}
 \ \Rightarrow \
 -iq\la x_4=\frac{x_{12}(x_{124}-i\la\ga Kx)}{i\la x},
\end{equation}
where we have used the periodic condition \eqn{pc} in rewriting the last equation.
Using \eqn{f3421} to eliminate $x_{124}$ in \eqn{C214k}, we finally have
\beqn\label{T4T1T2}
x_4=\frac{\left((-\al^2+\be^2)x_1x_2-ix\ga\la(\al x_1-\be x_2)\right)K}
{\la(\be x_1-\al x_2)}.
\eeqn

\section{Relation to $q$-Painlev\'e equations}
In this section, we identify $q$-Painlev\'e equations from the (1,1,1)-reduction of the 4-cube described in \S 4. 
Observing that the reduction collapses the 3-cubes $C_{312}$ and $C_{3124}$ in Figure \ref{red4cube} to two copies of a hexagon, which can be extended everywhere in the triangular lattice, we find the affine Weyl group $\widetilde{\mathcal W}\bigl(A_2^{(1)}\bigr)$. But the fourth direction relating the two 3-cubes provides us with an extra direction that leads to a lattice with Affine Weyl symmetry group $\widetilde{\mathcal W}\bigl((A_2+A_1)^{(1)}\bigr)$, which is the full symmetry group of the $A_5^{(1)}$-surface $q$-Painlev\'e equation. A sub-case of the (1,1,1)-reduction leads us to the symmetric version of this equation, which is often called the second $q$-discrete Painlev\'e equation (\qtw).

Define
\beqn\label{gt}
x(l,m,n,k)=(-1)^{\frac{1}{2}(l+m+n+k)}\la^{m+\frac{\be_1}{\ln q}} \om(l,m,n,k),
\eeqn
then the periodic condition $\ohlt x=-i\la x$ implies
\beqn\label{pcom}
\ohlt \om=\om.
\eeqn
On letting 
$K=\la^{-1}(q\la^2-1)$, $\be\al^{-1}=a_1$, $\al\ga^{-1}=q^{-1}a_0$ and $\ga\be^{-1}=a_2$,
our $q$-periodically reduced system on the asymmetric 4-cube is exactly the $\omega$-lattice constructed from the $\tau$-function frame work of $A_5^{(1)}$-surface $q$-Painlev\'e system \cite{jns:14}.
Results from \cite{jns:14} are reproduced in Appendix A to show that iterations of $\om$ in the $l,m,n$ and $k$ direction give rise to the affine Weyl  group $\widetilde{\mathcal W}\bigl((A_2+A_1)^{(1)}\bigr)$. 
\par Figure \ref{4cube om} shows the vertices of the 4-cube now relabeled in $\om$ variables. The quad-equations on the faces adjacent to $\om$ are:

\begin{minipage}{0.45\textwidth}
\begin{subequations}
\begin{align}\label{w12}
&\frac{\la\ohl\om}{\om}=\frac{\al\ol \om-\be\la\oh \om}{-\be \ol \om+\al\la\oh \om}\\\label{w13}
&\frac{\olt\om}{\om}=\frac{\al\ol \om-\ga\ot \om}{-\ga \ol \om+\al\ot \om}\\\label{w23}
&\frac{\la\oht\om}{\om}=\frac{\be\la\oh \om-\ga\ot \om}{-\ga\la \oh \om+\be\ot \om}
\end{align}
\end{subequations}
\end{minipage}\ \ \ \ \ 
\begin{minipage}{0.45\textwidth}
\begin{subequations}
\begin{align}\label{w14}
&\frac{\ool\om}{\om}=\frac{\al K}{\be}+\frac{\oo \om}{\ol \om}\\\label{w24}
&\frac{\ooh\om}{\om}=\frac{K}{q\la}+\frac{\oo \om}{q\la^2\oh \om}\\\label{w34}
&\frac{\oot\om}{\om}=\frac{K \ga}{\be}+\frac{\oo \om}{\ot \om}
\end{align}
\end{subequations}
\end{minipage}

The expression relating $\oo{}$, $\ol{}$\ and $\oh{}$ directions is given by transforming Equation \eqref{T4T1T2} to $\om$ variables:
\beqn\label{wT4T1T2}
\oo \om=\frac{\la\left((-\al^2+\be^2)\ol \om\oh\om-\ga\om(\al \ol\om-\be \la\oh\om)\right)K}
{\be(-1+q\la^2)(\be \ol\om-\al \oh\om)}.
\eeqn
\begin{figure}[ht!] 
\centering
\begin{tikzpicture}[scale=0.6]	

 	\foreach \s in {1}
	{
\draw (0+\s,\s+0.3) -- (0+\s,\s+5.3);
\draw (0+\s,\s+0.3) -- (6+\s,\s+0.3);
\draw (6+\s,\s+0.3) -- (6+\s,\s+5.3);
\draw (0+\s,\s+5.3) -- (6+\s,\s+5.3);

\draw (6+\s,\s+0.3) -- (8+\s,\s+1);
\draw (6+\s,\s+5.3) -- (8+\s,\s+6);
\draw (0+\s,\s+5.3) -- (2+\s,\s+6);
\draw (2+\s,\s+6) -- (8+\s,\s+6);
\draw (8+\s,\s+1) -- (8+\s,\s+6);

\draw (0+\s,\s+0.3) -- (2+\s,\s+1);
\draw (2+\s,\s+1) -- (2+\s,\s+6);
\draw (2+\s,\s+1) -- (8+\s,\s+1);
}
\foreach \r in {0.4}
 	{\foreach \s in {0.8}
	{
	\foreach \t in {0.6}
	{
	\draw (2+\s,2+\r) -- (2+\s,4.5+\r);
	\draw (2+\s,2+\r) -- (4.5+\s,2+\r);
	\draw (4.5+\s,2+\r) -- (4.5+\s,4.5+\r);
	\draw (2+\s,4.5+\r) -- (4.5+\s,4.5+\r);

	\draw (4.5+\s,2+\r) -- (5.5+\s,-\t+3+\r);
	\draw (4.5+\s,4.5+\r) -- (5.5+\s,-\t+5.5+\r);
	\draw (2+\s,4.5+\r) -- (3+\s,-\t+5.5+\r);
	\draw (3+\s,-\t+5.5+\r) -- (5.5+\s,-\t+5.5+\r);
	\draw (5.5+\s,-\t+3+\r) -- (5.5+\s,-\t+5.5+\r);
	\draw[very thick,blue] (2+\s,2+\r) -- (3+\s,-\t+3+\r);
	\draw[very thick,green] (3+\s,-\t+3+\r) -- (3+\s,-\t+5.5+\r);
	\draw[very thick,red] (3+\s,-\t+3+\r) -- (5.5+\s,-\t+3+\r);
	\filldraw [black]  (3+\s,-\t+3+\r) node [anchor=east] {$\om$} circle(0.5ex);
	\filldraw [black]  (2+\s,2+\r) node [anchor=east] {$\ot\om$} circle(0.5ex);
	\filldraw [black]  (3+\s,-\t+5.5+\r) node [anchor=east] {$\oh\om$} circle(0.5ex);
	\filldraw [black]  (5.5+\s,-\t+3+\r) node [anchor=west] {$\ol\om$} circle(0.5ex);
	
	\filldraw [black]  (2+\s,4.5+\r) node [anchor=east] {$\oht\om$} circle(0.5ex);
	\filldraw [black]  (5.5+\s,-\t+5.5+\r) node [anchor=west] {$\ohl\om$} circle(0.5ex);
	\filldraw [black]  (4.5+\s,2+\r) node [anchor=west] {$\olt\om$} circle(0.5ex);
	\filldraw [black]  (4.5+\s,4.5+\r) node [anchor=west] {$\ohlt\om$} circle(0.5ex);
	
	}
	}
}
\draw[dashed] (2+0.8,2+0.4) -- (0+1,1+0.3);
\draw[dashed] (2+0.8,4.5+0.4) -- (0+1,1+5.3);
\draw[dashed] (3+0.8,-0.6+5.5+0.4) -- (2+1,1+6);
\draw[very thick,yellow,dashed] (3+0.8,-0.6+3+0.4) -- (2+1,1+1);
\draw[dashed] (5.5+0.8,-0.6+3+0.4) -- (8+1,1+1);
\draw[dashed] (4.5+0.8,2+0.4) -- (6+1,1+0.3);
\draw[dashed] (4.5+0.8,4.5+0.4) -- (6+1,1+5.3);
\draw[dashed] (5.5+0.8,-0.6+5.5+0.4) -- (8+1,1+6);

        \filldraw [black]  (0+1,1+0.3) node [anchor=east] {$\oot\om$} circle(0.5ex);
	\filldraw [black]  (6+1,1+0.3) node [anchor=west] {$\oolt\om$} circle(0.5ex);
	\filldraw [black]  (0+1,1+5.3) node [anchor=east] {$\ooht\om$} circle(0.5ex);
	\filldraw [black]  (2+1,1+1) node [anchor=north] {$\oo\om$} circle(0.5ex);
	\filldraw [black]  (8+1,1+6) node [anchor=west] {$\oohl\om$} circle(0.5ex);
	\filldraw [black]  (8+1,1+1) node [anchor=west] {$\ool\om$} circle(0.5ex);
	\filldraw [black]  (6+1,1+5.3) node [anchor=west] {$\oohlt\om$} circle(0.5ex);
	\filldraw [black]  (2+1,1+6) node [anchor=east] {$\ooh\om$} circle(0.5ex);
\end{tikzpicture}
\caption{Figure \ref{two 4cubes} relabeled in $\omega$ coordinates.}\label{4cube om}
\end{figure}

\begin{figure}[ht!] 
\centering
  \begin{tikzpicture}
               \foreach \s in {1,2,3,4,5,6}
               \foreach \t in {5}
		{
\draw[dashed] (0+\s,0+\t) -- (0.5+\s, 0.866025403784+\t);
\draw[dashed] (1+\s,0+\t) -- (0.5+\s, 0.866025403784+\t);
\draw[dashed] (0.5+\s, 0.866025403784+\t) -- (-0.5+\s, 0.866025403784+\t);
\draw[dashed] (-0.5+\s, 0.866025403784+\t) -- (-1+\s,0+\t) ;
\draw[dashed] (-1+\s,0+\t) -- (\s,0+\t);
\draw[dashed] (-1+\s,0+\t) -- (-0.5+\s, -0.866025403784+\t) ;
\draw[dashed] (\s,0+\t) -- (-0.5+\s, -0.866025403784+\t);
\draw[dashed] (-0.5+\s, -0.866025403784+\t) -- (0.5+\s, -0.866025403784+\t) ;
\draw[dashed] (\s,0+\t) -- (0.5+\s, -0.866025403784+\t);
\draw[dashed] (0.5+\s, -0.866025403784+\t) -- (1+\s,0+\t);
		}
 		\foreach \s in {1,2,3,4,5,6}
		 \foreach \t in {5}
		{
\draw[dashed] (0.5+\s,2.59807621135+\t) -- (1+\s, 3.46410161514+\t);
\draw[dashed] (0+\s,1.73205080757+\t) -- (0.5+\s, 2.59807621135+\t);
\draw[dashed] (1+\s,1.73205080757+\t) -- (0.5+\s, 2.59807621135+\t);
\draw[dashed] (0.5+\s, 2.59807621135+\t) -- (-0.5+\s, 2.59807621135+\t);
\draw[dashed] (-0.5+\s, 2.59807621135+\t) -- (-1+\s,1.73205080757+\t) ;
\draw[dashed] (-1+\s,1.73205080757+\t) -- (\s,1.73205080757+\t);
\draw[dashed] (-1+\s,0+\t) -- (-1.5+\s, -0.866025403784+\t) ;
\draw[dashed] (\s,1.73205080757+\t) -- (-0.5+\s, 0.866025403784+\t);
\draw[dashed] (-0.5+\s, 0.866025403784+\t) -- (0.5+\s, 0.866025403784+\t) ;
\draw[dashed] (\s,1.73205080757+\t) -- (0.5+\s, 0.866025403784+\t);
\draw[dashed] (0.5+\s, 0.866025403784+\t) -- (1+\s,1.73205080757+\t);

\draw[very thick, red] (3.5, 0.866025403784+\t) -- (4.5, 0.866025403784+\t) ; 
\draw[very thick, green] (3.5, 0.866025403784+\t) -- (3, 1.73205080757+\t)  ;
\draw[very thick, blue] (3.5, 0.866025403784+\t) -- (3, 0+\t)  ;
\draw[thick] (3, 0+\t) -- (2.5, 0.866025403784+\t) ; 
\draw[thick] (2.5, 0.866025403784+\t) -- (3, 1.73205080757+\t)  ;
\draw[thick] (3, 1.73205080757+\t) -- (4, 1.73205080757+\t)  ;
\draw[thick] (4, 1.73205080757+\t) -- (4.5, 0.866025403784+\t);
\draw[thick] (4.5, 0.866025403784+\t) -- (4, 0+\t)  ;
\draw[thick] (4, 0+\t) -- (3, 0+\t)  ;

\filldraw [black]  (3.5, 0.866025403784+\t) node [anchor=south] {$\oo\om$} circle(0.1ex);
\filldraw [black]  (4.5, 0.866025403784+\t) node [anchor=south] {$\ool\om$} circle(0.1ex);
\filldraw [black]  (3, 1.73205080757+\t) node [anchor=south] {$\ooh\om$} circle(0.1ex);
\filldraw [black]  (3, 0+\t) node [anchor=south] {$\oot\om$} circle(0.1ex);
\filldraw [black]  (2.5, 0.866025403784+\t) node [anchor=south] {$\ooht\om$} circle(0.1ex);
\filldraw [black]  (4, 1.73205080757+\t) node [anchor=south] {$\oohl\om$} circle(0.1ex);
\filldraw [black]  (4, 0+\t) node [anchor=south] {$\oolt\om$} circle(0.1ex);

}
 		\foreach \s in {1,2,3,4,5,6}
		{
\draw[dashed] (0+\s,0) -- (0.5+\s, 0.866025403784);
\draw[dashed] (1+\s,0) -- (0.5+\s, 0.866025403784);
\draw[dashed] (0.5+\s, 0.866025403784) -- (-0.5+\s, 0.866025403784);
\draw[dashed] (-0.5+\s, 0.866025403784) -- (-1+\s,0) ;
\draw[dashed] (-1+\s,0) -- (\s,0);
\draw[dashed] (-1+\s,0) -- (-0.5+\s, -0.866025403784) ;
\draw[dashed] (\s,0) -- (-0.5+\s, -0.866025403784);
\draw[dashed] (-0.5+\s, -0.866025403784) -- (0.5+\s, -0.866025403784) ;
\draw[dashed] (\s,0) -- (0.5+\s, -0.866025403784);
\draw[dashed] (0.5+\s, -0.866025403784) -- (1+\s,0);
		}
 		\foreach \s in {1,2,3,4,5,6}
		{
\draw[dashed] (0.5+\s,2.59807621135) -- (1+\s, 3.46410161514);
\draw[dashed] (0+\s,1.73205080757) -- (0.5+\s, 2.59807621135);
\draw[dashed] (1+\s,1.73205080757) -- (0.5+\s, 2.59807621135);
\draw[dashed] (0.5+\s, 2.59807621135) -- (-0.5+\s, 2.59807621135);
\draw[dashed] (-0.5+\s, 2.59807621135) -- (-1+\s,1.73205080757) ;
\draw[dashed] (-1+\s,1.73205080757) -- (\s,1.73205080757);
\draw[dashed] (-1+\s,0) -- (-1.5+\s, -0.866025403784) ;
\draw[dashed] (\s,1.73205080757) -- (-0.5+\s, 0.866025403784);
\draw[dashed] (-0.5+\s, 0.866025403784) -- (0.5+\s, 0.866025403784) ;
\draw[dashed] (\s,1.73205080757) -- (0.5+\s, 0.866025403784);
\draw[dashed] (0.5+\s, 0.866025403784) -- (1+\s,1.73205080757);
		}
\draw[very thick, red] (3.5, 0.866025403784) -- (4.5, 0.866025403784) ; 
\draw[very thick, green] (3.5, 0.866025403784) -- (3, 1.73205080757)  ;
\draw[very thick, blue] (3.5, 0.866025403784) -- (3, 0)  ;
\draw[very thick, yellow]    (3.5, 0.866025403784) -- (3.5, 0.866025403784+5);
\draw[thick] (3, 0) -- (2.5, 0.866025403784) ; 
\draw[thick] (2.5, 0.866025403784) -- (3, 1.73205080757)  ;
\draw[thick] (3, 1.73205080757) -- (4, 1.73205080757)  ;
\draw[thick] (4, 1.73205080757) -- (4.5, 0.866025403784);
\draw[thick] (4.5, 0.866025403784) -- (4, 0)  ;
\draw[thick] (4, 0) -- (3, 0)  ;

\filldraw [black]  (3.5, 0.866025403784) node [anchor=south] {$\om$} circle(0.1ex);
\filldraw [black]  (4.5, 0.866025403784) node [anchor=south] {$\ol\om$} circle(0.1ex);
\filldraw [black]  (3, 1.73205080757) node [anchor=south] {$\oh\om$} circle(0.1ex);
\filldraw [black]  (3, 0) node [anchor=south] {$\ot\om$} circle(0.1ex);
\filldraw [black]  (2.5, 0.866025403784) node [anchor=south] {$\oht\om$} circle(0.1ex);
\filldraw [black]  (4, 1.73205080757) node [anchor=south] {$\ohl\om$} circle(0.1ex);
\filldraw [black]  (4, 0) node [anchor=south] {$\olt\om$} circle(0.1ex);
\filldraw [black]  (5, 1.73205080757) node [anchor=south] {$\ohhl\om$} circle(0.1ex);
\filldraw [black]  (5.5, 0.866025403784) node [anchor=south] {$\oll\om$} circle(0.1ex);

\end{tikzpicture}
\caption{ 2 Hexagons from the outer and inner 3-cubes on $(A_2+A_1)^{(1)}$ lattice}\label{a2b lattice}
\end{figure}
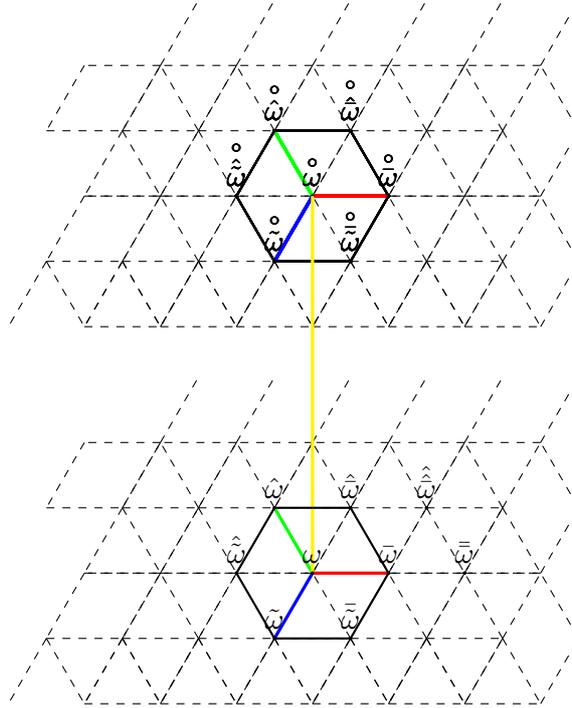

Using the fact $\ohlt \om=\om$, we see that the inner 3-cube in Figure \ref{4cube om} collapses to a hexagon, and similarly for the outer 3-cube using $\oohlt \om=\oo\om$. Iterations of these provide two copies of the triangular lattice drawn in Figure \ref{a2b lattice}.

Every vertex on the lower triangular lattice in Figure \ref{a2b lattice} 
can be calculated from the four initial conditions $\om,\;\ol\om,\;\oh\om,\;\ot\om$, using the three quad-equations \eqn{w12}--\eqn{w23}. (Vertices $\ohl \om, \olt\om$ and $\oht\om$ are calculated using \eqn{w12}, \eqn{w13} and \eqn{w23} respectively.) Equation \eqn{wT4T1T2} provides a link between the upper and lower triangular lattices. In this way, we find that the iteration in the $\oo{}$ direction (or $k$ direction) provides layers of triangular lattices. See Figure \eqref{a2b lattice}.

Without the (1,1,1)-periodic condition, the $\hIII$ system cannot be embedded into the layered triangular lattice which has affine Weyl symmetry $\widetilde{\mathcal W}\bigl((A_2+A_1)^{(1)}\bigr)$, which we call
here $(A_2+A_1)^{(1)}$ lattice. So the system provided above is a reduction of the $\hIII$ system on a 4-cube. To show that the reduced 4-cube system is indeed $q$-\Pa system on $A_5^{(1)}$-surface,
we define: 
\beqn\label{fg}
f=\frac{\ol \om}{\ohl \om},\;\; g=\la\frac{\ohl \om}{\om}
\eeqn
 and find their shifts in the $\ol{}$
direction, i.e., $\ol f={\oll \om}/{\ohll \om}$, $\ol g=\la\,{\ohll \om}/{\ol\om}$.
For $\ohll \om$, we use \eqn{w13}, shifted one step in $\ol{}$ and one step in $\oh{}$, along with
the periodic condition \eqn{pcom}. For $\oll\om$, we shift \eqn{w12} one step in $\ol{}$. These provide
\beqn\label{w112}
\ohll \om=\frac{q \al \om \ol\om+\ga\om\ohl\om}{\ga\ol\om+q\al \oh\om},
\;\;
\oll\om=\frac{\be\la\ol\om\ohl\om+q\al\la^2\ohl\om\ohll\om}{q\al\ol\om+\be\la\ohll\om}.
\eeqn
Thus we have
\beqn\label{qPIII}
\ol g=\frac{\la^2(1+ft)}{fg(f+t)},\;\;
\ol f=\frac{\la^2(1+a\ol g t)}{f\ol g (\ol g+at)},
\eeqn
where we have let 
\beqn\label{ta}
t=q\al/\ga,\;\;a=\ga/\be
\eeqn
and $t$ is the independent variable of the $q$-\Pa equation and $a$ is a parameter. The system \eqref{qPIII} is the third $q$-discrete Painlev\'e equation (\qth) \cite{sak:01}.

The triangular lattices in Figure \ref{a2b lattice} also provide a direct way of constructing the B\"acklund transformations of system \eqref{qPIII}. For translation in the $\oo{}$ direction for $g$ and $f$ we have $\oo g=q \la\oohl \om/\oo\om$ and
$\oo f=\ool \om/\oohl \om$. For $\oohl \om$ we use the equation on the face containing
$\ohlt\om$, $\ohl\om$, $\oohl\om$, $\oohlt\om$ and the periodic condition \eqn{pcom}. These lead to 
\beqn\label{wT124}
\oohl\om=\om \left(-\frac{K\ga}{q\be}+\frac{\oo\om}{\ohl\om}\right).
\eeqn
Using \eqn{wT4T1T2}, \eqn{wT124}, and \eqn{w14} we have:
\beqn\label{qPIIIBT}
\oo g=\frac{q(fgq+aft+a)}{g(fgq+aft+aq\la^2)},\;\;
\oo f=\frac{q(fg+aft\la^2+a\la^2)}{f(fgq+aft+a)},
\eeqn
which describe the \BT\ of $f$ and $g$, also known as the fourth $q$-discrete Painlev\'e equation (\qfo).
\subsection{(2,1) staircase reduction}

We now explain how different types of staircases taken on a two-dimensional square lattice are actually sub-cases of the geometric reduction on the $n$-cube.
We demonstrate this by investigating the only sub-case of the reduction considered in the previous section.

Let $l=n$, $\al=\ga$, (i.e., $\ol{}=\ot{}$\ ) then the (1,1,1)-periodicity $\ohlt\om=\om$ becomes the (2,1)-periodicity  
\beqn\label{pcom21}
\ohll\om=\om.
\eeqn System \eqn{w12}--\eqn{w34} reduces to one equation, i.e. Equation \eqn{w12} with the condition
${\oll\al}/{\al}={\oh \be}/{\be}=q^2$.
Using the same definition \eqn{fg} for $f$ and $g$, and \eqn{ta} for $t$ and $a$, we find
\beqn
\ol g=\frac{\la^2}{gf},\;\;
\ol f=\frac{g(at\ol g+1)}{\la(\ol g+at)},
\eeqn
which can be rewritten as a single equation for $g$
\beqn\label{qPII}
\oll g=\frac{\la^3(\ol g+at)}{\ol g g(at\ol g+1)}.
\eeqn
The resulting equation is the symmetric version of \qth \eqn{qPIII}, usually referred to as \qtw. This correspondence
between $(1,1,1)$ and $(2,1)$ reductions can be explained by projective reduction from the viewpoint of $\om$-lattice
\cite{jns:14}.

\section{Conclusion}
In this paper, we provided a new method called \lq\lq geometric reduction\rq\rq\ that relates ABS equations to discrete Painlev\'e equations. The method relies on the identification of an $n$-cube with the Voronoi cell of the root lattice of $\mathcal W\bigl(A_{n-1}\bigr)$.
As an example, we constructed a $q$-Painlev\'e equation from an asymmetric system based on the $\hIII$  and $H^6$ type equations on a 4-cube and provided the reduction to its symmetric form as its 2D sub-case. 
We also answered here a question posed at the end of \cite{O:14} whether a gauge transformation 
\eqn{gt} 
can be explained by the symmetry. It turns out to be the $\mathcal W\bigl(A_{1}^{(1)}\bigr)$ part of the
full symmetry of the \Pa system associated with rational surface of type $A_{5}^{(1)}$, i.e., its \BT.
\BT s\ of the discrete Painlev\'e equation arise as a natural by-product of our construction. Obtaining other structures related to integrability, such as Lax pairs, is also possible and will be reported in a separate paper. 
An interesting future direction is to extend our method to other types of discrete Painlev\'e equations classified by Sakai \cite{sak:01}, and moreover to understand its relations with
other types of higher dimensional integrable systems \cite{KNY:2002, Tsuda:09, Tsuda:10}. 

\noindent{\bf Acknowledgement.} 
The authors would like to express their sincere thanks to Drs. J. Atkinson, P. Kassotakis and P. McNamara
for inspiring and fruitful discussions.

\appendix
\section{The generators of $\widetilde{\mathcal W}\bigl((A_2+A_1)^{(1)}\bigr)$ and the triangular lattice}
The affine Weyl group $\widetilde{\mathcal W}\bigl((A_2+A_1)^{(1)}\bigr)$ is generated by $ s_0,s_1,s_2,\pi,w_0,w_1,r$, which are transformations of parameters and variables that satisfy the fundamental relations
\begin{align}
 &{s_i}^2=(s_is_{i+1})^3=\pi^3=1,\quad
 \pi s_i = s_{i+1}\pi,\quad
 (i\in\mathbb{Z}/3\mathbb{Z}),\\
 &{w_0}^2={w_1}^2=r^2=1,\quad
 rw_0=w_1r.
\end{align}
Here, the action of $\widetilde{W}(A_2^{(1)})=\langle s_0,s_1,s_2,\pi\rangle$ and 
that of $\widetilde{W}(A_1^{(1)})=\langle w_0,w_1,r\rangle$ commute. Full details of how to construct this affine Weyl group from the $\om$-lattice can be found in \cite{jns:14}, with the identification $\om=\om_0$. Here we point out how this construction can be related to the triangular lattice shown in Figure \ref{a2b lattice}.

Define the translations $T_i~(i=1,2,3,4)$ by
\begin{equation}
 T_1=\pi s_2s_1,\quad
 T_2=\pi s_0s_2,\quad
 T_3=\pi s_1s_0,\quad
 T_4=rw_0,
\end{equation}
where $T_i$ ($i=1, 2, 3$) are translations of $\widetilde{W}(A_2^{(1)})$ and $T_4$
is a translation of $\widetilde{W}(A_1^{(1)})$.
We connect these generators to the triangular lattice in Figure \ref{a2b lattice} by identifying
\begin{align*}
 &T_1:(a_0,a_1,a_2,\la)\to(qa_0,q^{-1}a_1,a_2,\la) \hspace{5mm}  \Leftrightarrow\hspace{5mm} \ol{}\ :\ l\mapsto l+1,\\
 &T_2:(a_0,a_1,a_2,\la)\to(a_0,qa_1,q^{-1}a_2,\la) \hspace{5mm}  \Leftrightarrow\hspace{5mm}
 \oh{}: \ m\mapsto m+1,\\
 &T_3:(a_0,a_1,a_2,\la)\to(q^{-1}a_0,a_1,qa_2,\la) \hspace{5mm}  \Leftrightarrow\hspace{5mm}
 \ot{}: \ n\mapsto n+1,\\
 &T_4:(a_0,a_1,a_2,\la)\to(a_0,a_1,a_2,q\la) \hspace{11mm}  \Leftrightarrow\hspace{5mm}
 \oo{}: \ k\mapsto k+1.
\end{align*}
The periodicity condition $\ohlt\om=\om$ corresponds to the relation $T_1T_2T_3=1$ of 
$\widetilde{W}(A_2^{(1)})$.


\end{document}